\documentstyle[aps]{revtex}
\begin{document}
\draft
\twocolumn

\title{Restoration of Overlap Functions
                     and
        Spectroscopic Factors in Nuclei}

\author{M. V. Stoitsov, S. S. Dimitrova and A. N. Antonov }

\address{Institute for Nuclear Research and Nuclear Energy, \\
        Bulgarian Academy of Sciences, \\
        Sofia-1784, Bulgaria \\~~\\~~}

\maketitle

\begin{abstract}
An asymptotic restoration procedure is applied for
analyzing bound--state overlap functions, separation energies and
single--nucleon spectroscopic factors by means of a model one--body
density matrix emerging from the Jastrow correlation method in its
lowest order approximation for  $^{16}O$ and $^{40}Ca$ nuclei .
Comparison is made with available experimental data and mean--field and
natural orbital representation results.
\end{abstract}
\pacs{PACS number(s):~21.10.Gv,21.60.Cs,21.60.Jz}

Both the single--nucleon spectroscopic factors and the overlap
functions have attracted much attention in questions concerning the
interpretation of recent $(e,e\prime p)$, $(d,^{3}He)$ and $(\gamma,p)$
experimental data (e.g.
\cite{r1,r2,r3,r4,r5,r6,r7,r8,r9,r10,r11,r12,r13,r14,r15}.
  The growing interest is motivated in principle by the possibility to
 clarify the limitation of the nuclear
mean--field picture.
For instance, the relatively low values of the spectroscopic factors deduced
from these experiments show clearly the importance of the short-range
correlation effects in nuclei and the necessity of detailed investigations
of the  high--momentum components of the nucleon wave function which
cannot be included within the mean--field approximation
 \cite{r16,r17,r18,r19,r20}.

The underlying relationship between the differential cross--section and
the structure of the nuclear wave function is  often empirically analyzed
 using the plane--wave impulse approximation (PWIA). For instance, in this
 approximation the $(e,e\prime p)$--reaction cross--section for a transition
 to a specific state  with quantum numbers $\alpha$ in
the residual nucleus has the following form (see e.g.
\cite{r12,r22})
\begin{equation}
\sigma_{(e,e\prime p)}
{}~\equiv~{ {d^{5} \sigma} \over {d\Omega_{e\prime}d\Omega_{p}dE_{p}}}~ =
{}~{\cal{K}}~ \sigma_{ep}{\mid \phi_{\alpha}({\bf k}) \mid^{2} }~.
\label{dwba}
\end{equation}
The first term $\cal{K}$ is kinematical factor , $\sigma_{ep}$ is the
off--shell electron--proton scattering cross--section \cite{r23} and
the  nuclear structure component $\mid\phi_{\alpha}({\bf k})\mid^{2}$  is
 the squared Fourier transform of the overlap function between the
 ground state of the target nucleus $\Psi^{(A)}$ and the final state of
the residual nucleus $\Psi^{(A-1)}$ \cite{r24,r25,r16}:
\begin{equation}
\phi_{\alpha}({\bf r}) = \langle
\Psi^{(A-1)}_{\alpha} \mid a({\bf r}) \mid \Psi^{(A)} \rangle ~,
\label{ovf}
\end{equation}
$a({\bf r})$ being an  annihilation
 operator for a nucleon with spatial coordinate ${\bf r}$ (spin
and isospin coordinates are not put in evidence).  The overlap functions
(\ref{ovf}) are not orthonormalized.  Their norm defines the
spectroscopic factor of the level $\alpha$
\begin{equation}
S_{\alpha}  =  \langle \phi_{\alpha} \mid \phi_{\alpha} \rangle
\label{spf}
\end{equation}
and the normalized overlap function
\begin{equation}
\tilde{\phi}_{\alpha}({\bf r})  =
S_{\alpha}^{-1/2} ~\phi_{\alpha}({\bf r})~.
\label{novf}
\end{equation}
Usually, $\tilde{\phi}_{\alpha}({\bf r})$ is calculated from an
empirical Saxon--Woods potential with a distinct potential radius for
each separate transition $\alpha$.  Quantitative estimates are then
deduced by fitting both the potential radius and the spectroscopic
factor $S_{\alpha}$ in order to obtain a good agreement between the
experimentally measured cross sections $\left( {d\sigma} / {d\Omega}
\right)_{exp}$ and those predicted from appropriate calculations
for the reaction process.

The full theoretical description of the experiments mentioned above has
many components. We should like to mention among them the proper
account of the reaction mechanism, of the distortion effects (including
the distortion due to the final state interaction), of the meson
exchange currents contributions, the study of the A-dependence and
others (see e.g. \cite{r22,r12}). Obviously, however, the theoretical
estimates of the overlap functions
(\ref{ovf}) and the spectroscopic factors (\ref{spf}) are of crucial
importance for the adequate evaluation of recent $(e,e\prime p)$,
$(d,^{3}He)$ and $(\gamma,p)$ experiments.  The problem is that the
normalized overlap functions (\ref{novf}) cannot be identified with a
phenomenological shell--model single--particle wave function especially
for energies
farther from the Fermi energy, sometimes even within the valence shell
\cite{r16}. It is known that generally, the independent--particle shell
model cannot explain the fragmentation or spreading of the hole strength
\cite{r26,r22,r20}. This is because, due to the residual interaction,
the hole state of the target nucleus is not an eigenstate of the
$(A-1)$--nucleon system and its strength is distributed over several
states of the final system.  Possible modifications going beyond the
uncorrelated shell--model approximation quickly become rather involved.

Recently, it has been shown \cite{r27} that the knowledge of the
ground--state one--body density matrix of the target nucleus is
sufficient to determine, at least in principle, the overlap functions,
spectroscopic factors and separation energies of the bound
$(A-1)$--particle eigenstates.  The aim of the present paper is to apply
the procedure suggested in \cite{r27} using the model one--body density
matrix \cite{r28,r29,r30} in which the short--range correlation terms of
the Jastrow correlation method are taken into account.
The resulting quantitative estimates allow
us to make instructive conclusions for the properties of the overlap
functions in comparison with the associated shell--model orbitals and
the natural orbitals \cite{r31} which are of frequent interest in this
context \cite{r16,r19,r20}.

The exact one--body density matrix associated with the ground state
$\Psi^{(A)}$ of the target nucleus with $A$ nucleons is defined as
\begin{equation}
\rho({\bf r},{\bf r}\prime)  =
\langle \Psi^{(A)} \mid a^\dagger({\bf r})~ a({\bf r}\prime)
\mid \Psi^{(A)} \rangle ~.
\label{odm}
\end{equation}
Inserting the complete set of eigenstates $\Psi_{\alpha}^{(A-1)}$ of the
residual $(A-1)$--nucleus, eq.~(\ref{odm}) reads
\begin{equation}
\rho({\bf r},{\bf r}\prime) = \displaystyle{
\sum_{\alpha}~ \phi^\ast_{\alpha}({\bf r})~ \phi_{\alpha}({\bf r}\prime)
= \sum_{\alpha}~ S_{\alpha}~ \tilde{\phi}_{\alpha}^{\ast}({\bf r})~
\tilde{\phi}_{\alpha}({\bf r}\prime) } ~,
\label{odms}
\end{equation}
where $\phi_{\alpha}$ and $\tilde{\phi}_{\alpha}$ are the overlap
functions (\ref{ovf}) and (\ref{novf}), respectively, $S_{\alpha}$ is
the spectroscopic factor (\ref{spf}) and the summation implicitly
includes also the continuum states associated with all scattering
channels of the $(A-1)$ system.

The one--body density matrix has a quite similar form in its natural
orbital representation \cite{r31}
\begin{equation}
\rho({\bf r},{\bf r}\prime)  =  \displaystyle{
\sum_{\alpha}~ N_{\alpha}~ \psi_{\alpha}^{\ast}({\bf r}) ~
\psi_{\alpha}({\bf r}\prime) },
\label{odmn}
\end{equation}
where the natural orbitals $\psi_{\alpha}$ are defined as a complete
orthonormal set of functions which diagonalize the one-body density
matrix (\ref{odm}) with eigenvalues $N_{\alpha}$ called natural
occupation numbers.  Properties of the overlap functions and the natural
orbitals are considered for various many-body systems, such as atomic
nuclei \cite{r16,r28,r29,r19,r20,r32,r33,r34} and liquid drops of
$^{3}He$ \cite{r35}.

In the case of a spherical ground state $\Psi^{(A)}$ with $0^{+}$
angular momentum and parity, each eigenstate $\Psi^{(A-1)}_\alpha$ is
characterized by the `single--particle' quantum numbers $ljm$, i.e.,
$\alpha \equiv nljm$ with $n$ being the number of the $ljm$--state.
 The overlap functions and the natural orbitals then factorize
into radial and spin--angular parts
\begin{equation}
\begin{array}{l}
\phi_\alpha ({\bf r}) = \phi_{nlj}(r)~Y_{ljm}(\Omega,\sigma), \\
{}~~~ \\
\psi_\alpha ({\bf r}) = \psi_{nlj}(r)~Y_{ljm}(\Omega,\sigma), \\
\end{array}
\label{symm}
\end{equation}
where
$Y_{ljm}(\Omega,\sigma) = [Y_l(\Omega) \times \chi^{1/2}(\sigma)]^j_m$
and $\sigma$ is the
spin variable. Using eq.~(\ref{symm}), the one--body density matrix
reads
\begin{equation}
\rho({\bf r},{\bf r}\prime)  =  \displaystyle{
\sum_{lj}~\rho_{lj}(r,r\prime)~\sum_{m
\sigma}~Y^{\ast}_{ljm}(\Omega,\sigma)~ Y_{lj m}(\Omega\prime,\sigma) }~.
\label{rhol}
\end{equation}
Due to the spherical symmetry $S_{nlj}$, $N_{nlj}$ and the radial
contributions $\rho_{lj}(r,r\prime)$ entering the one--body density
matrix (\ref{rhol}) do not depend on the magnetic quantum number $m$.
{}From eqs.~(\ref{odms}) and (\ref{odmn}) then follows that in each
$lj$--subspace the spectroscopic factor $S_{nlj}$ is smaller than the
largest natural occupation number $N_{nlj}^{\max}$ with the same $lj$,
i.e.,
\begin{equation}
S_{nlj} ~ \le~ N_{nlj}^{\max}~.
\label{max}
\end{equation}

The procedure of  \cite{r27} is based on the generally accepted
asymptotic
behavior of the neutron overlap functions associated with the bound
states of the $(A-1)$ system
\begin{equation}
\phi_{nlj}(r)~\rightarrow~C_{nlj}~\exp(-k_{nlj}~r)/r~,
\label{asf}
\end{equation}
where
\begin{equation}
k_{nlj} = \hbar^{-1} \sqrt{ 2m(E^{(A-1)}_{nlj}~-~E^{(A)}_{0}) }
\label{ask}
\end{equation}
depends on the separation energy $\epsilon_\alpha =
E^{(A-1)}_{nlj}~-~E^{(A)}_{0}$.  Obviously, the higher excited states
have faster decay.  Therefore, for large values
$r\prime = a \rightarrow \infty$,
eqs.~(\ref{odms}) and (\ref{rhol}) lead to the asymptotic
relation
\begin{equation}
\rho_{lj}(r,a)~ \rightarrow ~\phi_{n_{0}lj}(r) ~\times ~C_{n_{0}lj}~
\exp(-k_{n_{0}lj}a)/a~,
\label{asr}
\end{equation}
where $\phi_{n_{0}lj}(r)$ is the radial part of the lowest overlap
function in the $lj$ subspace considered.  The unknown constant
$C_{n_{0}lj}$ can be derived from the asymptotic behavior of the
diagonal part $\rho_{lj}(r,r)$ since
\begin{equation}
\rho_{lj}(a,a)~ \rightarrow ~\mid C_{n_{0}lj}\mid^2~
\exp( - 2 k_{n_{0}lj}a)/a^2~.
\label{asd}
\end{equation}

By means of eqs.~(\ref{asr}) and (\ref{asd}) one can derive the lowest
(bound state) overlap function  with quantum numbers $lj$
and radial part
\begin{equation}
\phi_{n_{0}lj}(r)  = { {\rho_{lj}(r,a)} \over
{C_{n_{0}lj}~\exp(-k_{n_{0}lj}a)/a}}~
\label{ovf0}
\end{equation}
as well as the associated separation energy
\begin{equation}
\epsilon_{n_{0}lj}  = { {\hbar^2~k^2_{n_{0}lj} } / {2m}}
\label{k0}
\end{equation}
and the spectroscopic factor
\begin{equation}
S_{n_{0}lj} =  \langle \phi_{n_{0}lj} \mid \phi_{n_{0}lj} \rangle ~.
\label{s0}
\end{equation}
One can repeat the above procedure for the second bound state with
the same multipolarity (if
it exists) after subtracting the contribution of the lowest eigenstate.
The result is
\begin{equation}
\phi_{n_{1}lj}(r)  = {
{ \rho_{lj}(r,a) - \phi_{n_{0}lj}(r)~\phi_{n_{0}lj}(a) } \over
{ C_{n_{1}lj}~\exp (-k_{n_{1}lj}a)/a}  }~
\label{ovf1}
\end{equation}
and expressions similar to (\ref{k0}) and (\ref{s0}) for the separation
energy $\epsilon_{n_{1}lj}$ and the spectroscopic factor $S_{n_{1}lj}$,
respectively.  The restoration procedure can be continued and one is
able to analyze all bound states of the $(A-1)$--particle system once
the one--body density matrix of the $A$--particle ground state is known.
In the case of proton bound--states some modifications due to the
Coulomb asymptotic behavior of the overlap functions have to be taken
into account.

The present calculations of the bound--state overlap functions,
separation energies and spectroscopic factors have been performed
applying the recipe (\ref{asr}) -- (\ref{ovf1}) to a model one--body
matrix \cite{r28,r29,r30} obtained within the Jastrow correlation
method in its low--order approximation for  $^{16}O$ and
$^{40}Ca$ nuclei .  The model is based on harmonic oscillator single Slater
determinant and Gaussian--like state--independent correlation factor.
Although the resulting density matrix has a simple analytical form, it
is physically significant that the short-range correlations are
 incorporated in it to a large extent. In addition, its natural orbital
representation is well investigated \cite{r28,r29,r30}.

Here we should like to mention that the procedure suggested in
\cite{r27} requires accurate values of the one--body density matrix at
large $r$. In principle this limits the practical application of the
method. In our opinion, however, the analytical expressions of the
one--body density matrix obtained within some correlation methods give a
basis for an easier numerical application of the procedure. This is the
case in the present work. Using the model one--body density matrix from
\cite{r28,r29,r30} we have to avoid, however, another difficulty arising
from its Gaussian asymptotic behavior. This is achieved by applying the
recipe   (\ref{asr}) -- (\ref{ovf1}) not for a single asymptotic
point $a$ but within an asymptotic region ($a_L,a_U$) bracketing the
point $a$ and sustained after the point where the diagonal element
$\rho_{lj}(r,r)$ is less than 10 percent from its maximal value.  We
are looking for such a radial contribution $\rho_{n_{0}lj}(r,r\prime) =
\phi_{n_{0}lj}(r)~\phi_{n_{0}lj}(r\prime)$ whose diagonal part is less
than or equal to $\rho_{lj}(r,r)$ at each point $r$ and which minimizes
the trace
$Tr \left[ (\rho_{lj} - \rho_{n_{0}lj})^2 \right]$.  Then $a_L$, $a$ and
$a_U$ as well as the unknown $C_{n_{0}lj}$ and $k_{n_{0}lj}$ are
uniquely determined by the requirement that the overlap function
(\ref{ovf0})
 satisfies eqs.~(\ref{asr}) and (\ref{asd}) simultaneously with minimal
least--squared deviation within the region ($a_L,a_U$). We should
mention that the procedure suggested is not the unique one, but this
problem does not exist when a realistic one-body density matrix with a
correct exponential asymptotic behavior is considered.

We have performed the above numerical procedure separately for each set of
quantum numbers $nl$ (the model does not split the states with respect
to $j = l \pm 1/2$).  It leads to predictions for the neutron separation
energies $\epsilon_\alpha$, spectroscopic factors $S_\alpha$ and the
overlap functions $\tilde{\phi}_\alpha$ which are given in Table~I and
Fig.~1, respectively.


It is seen from Table~I that the calculated separation energies
$\epsilon_\alpha$ are in acceptable agreement with the self--consistent
Hartree-Fock (HF) results \cite{r36} and the available experimental
data. The calculated spectroscopic factors $S_\alpha$, however, differ
significantly from their mean--field values.  Due to the short--range
correlations a depletion of the states below and a filling of the states
above the Fermi level results.  At the same time the calculated values
of $S_\alpha$ are consistent with experimentally deduced spectroscopic
factors \cite{r6}.  In general, the values of $S_\alpha$ emerging from
the present restoration procedure are larger than the experimental ones.
This fact is most probably related to the crude approximation for the
density matrix used.  The reason is the same for the larger value of
$S_{2s1/2}$ in comparison with the spectroscopic factor of the lower
$1s$-state in $^{40}Ca$.

In Table~I we compare $S_\alpha$ also with the natural occupation
numbers $N_\alpha$ derived after diagonalyzing the same model one--body
density matrix \cite{r28,r29,r30}. The comparison shows that our
numerical procedure satisfies the general requirement (\ref{max}).
The trend of the calculated spectroscopic factors $S_{\alpha}$ follows
that one of the natural occupation numbers.  This result becomes more
transparent realizing that the overlap functions $\tilde{\phi}_\alpha$
are rather close to the natural orbitals $\psi_\alpha$ as it is seen
from Fig.~1.

{}From Fig.~1 it can be also seen that all three functions, the overlap,
mean--field and natural orbital wave functions, are quite similar for
the hole states in nuclei.  This justifies the use of shell--model
orbitals instead of overlap functions within PWIA calculations
(\ref{dwba}) for such kind of nuclear states.  This approximation,
however, is no longer valid for the particle nuclear states where the
mean--field wave--functions significantly differ from the overlap
functions (see the $1f$-state in Fig.~1).  The latter take some
intermediate position between the natural orbitals and the HF
 wave--functions.
It should be stressed that our model one--body density matrix is
completely different from the Hartree--Fock (HF) one. It has
been demonstrated in \cite{r28,r29,r35} that due to the short--range
correlations the correlated particle--state orbitals are much more
localized than the particle--state mean--field single--particle wave
functions. This is the reason why the HF orbitals associated with
the particle--states go further out than the overlap functions (or
the natural orbitals) associated with the correlated one--body
density matrix. The place of the correlated particle-state
asymptotic region is affected by the SRC while the HF
particle-state orbitals have a larger spread although the orbit
is more strongly bound.

  The instructive conclusion is that neither natural
 orbitals nor shell--model wave--function can be used as particle--state
 overlap functions in the theoretical analysis of  the experimental
data.  The present restoration procedure gives a possible solution of
the problem  if it is applied to some realistic ground--state one--body
density  matrices.


Concluding, we have demonstrated in this paper the possibility for
restoring the separation energies, spectroscopic factors and overlap
functions for bound $(A-1)$-particle eigenstates on the basis of
 the ground--state one--body density matrix of the target $A$-particle
system.  Although we have used quite crude approximation for the
one--body density matrix \cite{r28,r29,r30}, the asymptotic restoring
procedure \cite{r27} leads to acceptable quantitative results.  Thus
one obtains a method for estimating such important quantities as
spectroscopic factors and overlap functions which is supplemental to the
more involved approaches \cite{r16}.  For this purpose, one has simply
to apply the present restoring procedure to more sophisticated one--body
density matrices as for example the ones emerging from
Brueckner--Hartree--Fock \cite{r40,r41}, variational Monte--Carlo
\cite{r35,r42} or simplectic model calculations \cite{r43}.
The resulting  bound--state spectroscopic factors and overlap functions
will have more realistic properties and can be used for reliable
description of the characteristics of the $(e,e\prime p)$, $(d,^{3}He)$ ,
$(\gamma,p)$ and other one--nucleon removal nuclear processes.

\acknowledgments

This work is supported in part by the Contracts $\Phi-406$ and
$\Phi-527$ with the Bulgarian National Science Foundation.

\baselineskip = 14pt
\bibliographystyle{unsrt}

\newpage
\narrowtext

\begin{table}
\caption{Separation energies $\epsilon_\alpha$ and spectroscopic factors
$S_\alpha$ calculated on the base of the one--body density matrix
\protect\cite{r28,r29,r30} for $^{16}O$ and $^{40}Ca$.  Comparison is
made with the Hartree--Fock (HF) single--particle energies (set SkI from
\protect\cite{r36}), natural occupation numbers $N_\alpha$
\protect\cite{r29} and experimental data (EXP).  }
\begin{tabular}{cccccccccccc}
& nl &~~&$\epsilon_\alpha^{HF}$& $\epsilon_\alpha$  & EXP & ~~~~ &
$S_\alpha^{HF}$ &
$S_\alpha$ & $N_\alpha$ & EXP \cr
\tableline
{}~~\cr
$^{16}O$ &1s &~&32.96 &35.82 & 47.$^{a)}$ &
{}~~ & 1 & 0.940 & 0.95~ & ~~~~ \cr
         &1p &~&20.81 &17.48 & 21.8$^{b)}$      &
{}~~ & 1 & 0.953 & 0.965 & ~~~~ \cr
         &1d &~&~5.31 &12.76 & ~4.14$^{b)}$     &
{}~~ & 0 & 0.004 &
0.006 & ~~~~ \cr~~\cr
$^{40}Ca$&1s &~&41.04 &32.85 & 56.$^{a)}$  &
 ~~ & 1 & 0.763  & 0.89~ & 0.75$^{d)}$ \cr
         &1p &~&32.17 &29.54 & 40.$^{a)}$   &
{}~~ & 1 & 0.89~ & 0.938 & 0.72$^{d)}$ \cr
         &1d &~&22.16 &24.75 & 22.38$^{c)}$     &
{}~~ & 1 & 0.907 & 0.946 & 0.74$^{d)}$ \cr
         &2s &~&15.67 &13.07 & 18.2$^{c)}$ &
{}~~ & 1 & 0.953 & 0.958 & 0.64$^{d)}$ \cr
         &1f &~&11.25 &~8.69 & ~8.36$^{c)}$ &
{}~~ & 0 & 0.01~ & 0.013 & 0.11$^{d)}$ \cr
\end{tabular}
\label{table1}
\tablenotetext{The energies are in MeV and only the states with $j+1/2$
are displayed.  Experimental data are taken:       $^{a)}$ from
\protect\cite{r37}, $^{b)}$ from \protect\cite{r38}, $^{c)}$ from
\protect\cite{r39}, $^{d)}$ from \protect\cite{r6}.  }
\end{table}


\begin{figure}
\caption[]{ Overlap functions (solid line), self-consistent Hartree-Fock
single--particle wave functions (dot-dashed line) and natural orbitals
(dashed line) for the nucleus $^{40}Ca$.  }
\label{ff1}
\end{figure}

\end{document}